\documentclass[twocolumn,showpacs]{revtex4}
\usepackage{times,xspace}
\usepackage{amsbsy,amssymb,amsmath,bm}
\usepackage{graphicx,color,epsfig,rotate}
\usepackage{fancyheadings}

\def\bbbc{{\mathchoice {\setbox0=\hbox{$\displaystyle\rm C$}\hbox{\hbox
to0pt{\kern0.4\wd0\vrule height0.9\ht0\hss}\box0}}
{\setbox0=\hbox{$\textstyle\rm C$}\hbox{\hbox
to0pt{\kern0.4\wd0\vrule height0.9\ht0\hss}\box0}}
{\setbox0=\hbox{$\scriptstyle\rm C$}\hbox{\hbox
to0pt{\kern0.4\wd0\vrule height0.9\ht0\hss}\box0}}
{\setbox0=\hbox{$\scriptscriptstyle\rm C$}\hbox{\hbox
to0pt{\kern0.4\wd0\vrule height0.9\ht0\hss}\box0}}}}

\newcommand{\ignore}[1]{}
\newcommand{\mComment}[1]{}
\newcommand{\gComment}[1]{}
\newcommand{\jComment}[1]{}
\newcommand{\rComment}[1]{}
\newcommand{\lComment}[1]{}

\renewcommand{\mComment}[1]{\textcolor{blue}{Manny: #1}}
\renewcommand{\gComment}[1]{\textcolor{red}{Gerardo: #1}}
\renewcommand{\jComment}[1]{\textcolor{green}{Jim: #1}}
\renewcommand{\rComment}[1]{\textcolor{magenta}{Ray: #1}}
\renewcommand{\lComment}[1]{\textcolor{purple}{Rolando: #1}}

\begin{document}
\title{Electronic Ferroelectricity in the Falicov-Kimball Model}
\author{C. D. Batista}
\affiliation{Theoretical Division, 
Los Alamos National Laboratory, Los Alamos, NM 87545}

\date{Received \today }

\begin{abstract}
I show that a spontaneous electric polarization 
exists in the solution of the Falicov-Kimball model 
by mapping the strong coupling limit of this Hamiltonian into an $xxz$ spin $1/2$ 
model with a magnetic field. In this way, I determine the 
phase diagram of the strongly interacting model and show the existence 
a transition to a mixed valence regime containing  two phases: 
an orbitally ordered state and a Bose-Einstein condensation of excitons with a built-in 
electric polarization. 
\end{abstract}


\maketitle

The Falicov-Kimball \cite{Falicov} model (FKM) was 
introduced to explain semiconductor-metal transitions and
has been extensively used to describe valence transitions 
in heavy fermion compounds. Its original version 
contains a dispersive band of itinerant $d$ electrons interacting
with localized $f$ orbitals via an on-site Coulomb interaction.
If hybridization between both bands is included, the $f$ charge occupation 
is no longer a good quantum number, and it is possible to build 
coherence between the $d$ electrons and the $f$ holes. Based on a
mean field solution of the FKM with a hybridization term,
Portengen {\it et al} \cite{Portengen} proposed that this coherence gives rise 
to a spontaneous electric polarization associated with a Bose-Einstein 
condensate (BEC) of $d-f$ excitons.  

Ferroelectrics are of considerable theoretical and technological interest because
of their highly unusual properties \cite{Kittel}.
The ferroelectric (FE) transitions have traditionally been considered as a
subgroup of the structural phase transitions. As in the 
case of superconductivity, the existence of ferroelectrics based
on a purely electronic mechanism would provide a new set of
physical properties \cite{Portengen} and technological applications; 
for instance, it would open the possibility of 
controlling optical properties with magnetic fields.

The proposal of electronic ferroelectricity in the FKM
was recently tested theoretically using different techniques.
An analytical calculation in infinite dimension for
the weak coupling limit \cite{Czycholl} did not confirm its
existence. By using numerical methods to solve finite-size
chains, Farka\v sovsk\'y \cite{Farka0} arrived at the same 
conclusion for the intermediate and the strong coupling regimes. 
Recently, Zlati\'c {\it et al} \cite{Zlatic} calculated the spontaneous polarization 
in the FKM from its exact solution  
in infinite dimensions. They found that the spontaneous hybridization
susceptibility diverges when the temperature goes to zero, indicating
a possible non-zero polarization of the ground state. 

Hybridization between the bands however is not the 
only way to develop $d-f$ coherence. An 
$f-f$ hopping also induces it.
Furthermore, I will show that in the mixed valence regime,
the ground state of the FKM with $f-f$ hopping 
is either an orbitally-ordered (chessboard) state or a BEC 
of electron-hole pairs (excitons). 
In particular, the condensate  has a built-in 
spontaneous FE or antiferroelectric (AFE) polarization 
induced by a pure electronic mechanism. The FE or AFE character of the 
solution depends on the relative sign of the $d-d$ and the $f-f$ hoppings.

I obtain the complete phase diagram of the extended FKM by mapping 
the original Hamiltonian into an effective spin model. The mapping is 
exact in the strong coupling limit and the resulting spin Hamiltonian is
a spin 1/2 $xxz$ model with an applied magnetic field along the ${\bf \hat z}$ axis. This 
spin model is exactly solvable in one dimension and its phase diagram 
has been determined very accurately for two dimensional ($D=2$) systems 
(the exact solution has been numerically obtained for a $96 \times 96$
square lattice \cite{Schmid}). In this way, the results obtained in this
paper prove that the phase diagram of the extended FKM 
contains FE and AFE phases induced by an electronic mechanism. 

 
Recently, high dielectric constants were observed in oxides of the 
type ACu$_3$Ti$_4$O$_{12}$. In particular, the largest dielectric constant ever 
observed is exhibited by CaCu$_3$Ti$_4$O$_{12}$ ($\epsilon_0 \simeq 80000$ for 
single-crystal samples at room temperature) \cite{Subramanian,Ramirez,Homes}. 
In addition, high resolution x-ray and neutron powder diffraction measurements 
of CaCu$_3$Ti$_4$O$_{12}$ rule out a conventional FE structural phase transition.
The electronic ferroelectricity proposed by Portengen {\it et al} \cite{Portengen}
was also ruled out due to the large value of the optical gap ($\Delta \sim 1.5 eV$)
\cite{He}. The strong coupling theory introduced in this paper shows that the excitons 
condense in the presence of a large gap. 

I will consider an extended FKM for spinless fermions on a $D$-dimensional
hypercubic lattice:  
\begin{eqnarray}
H &=& \epsilon_d \sum_{\bf i} n^d_{\bf i} + \epsilon_f \sum_{\bf i} n^f_{\bf i}
+t_d \sum_{\langle \bf i,j \rangle} d^{\dagger}_{\bf i} d^{\;}_{\bf j}
\nonumber \\
&+& U^{fd} \sum_{\bf i} n^d_{\bf i} n^f_{\bf i} 
+t_f \sum_{\langle {\bf i,j} \rangle} f^{\dagger}_{\bf i} f^{\;}_{\bf j},
\end{eqnarray}
where $n^d_{\bf i}= d^{\dagger}_{\bf i} d^{\;}_{\bf i}$, and
$n^f_{\bf i}= f^{\dagger}_{\bf i} f^{\;}_{\bf i}$ are the occupation 
numbers of each orbital.
For historical reasons, I denote the orbitals by $f$ and $d$ but 
in general they can represent any pair of atomic orbitals with different parity. For 
instance, in CaCu$_3$Ti$_4$O$_{12}$ there are 
$d$ states which are hybridized with $p$ bands \cite{He}. This fact is
another important motivation to include a non-zero $t_f$.

The Hamiltonian $H$ can be rewritten as an asymmetric Hubbard 
model if the orbital flavor is represented by a pseudospin variable. 
A spin 1/2 is required to describe the two orbitals (flavors) on each site:
\begin{eqnarray}
c^{\dagger}_{\bf i \uparrow}&=& d^{\dagger}_{\bf i} \;\;\;\; 
c^{\;}_{\bf i \uparrow} = d^{\;}_{\bf i},
\nonumber \\
c^{\dagger}_{\bf i \downarrow}&=& f^{\dagger}_{\bf i} \;\;\;\; 
c^{\;}_{\bf i \downarrow} = f^{\;}_{\bf i},
\label{dic}
\end{eqnarray}
where the pseudospin generators are:
\begin{eqnarray}
\tau^{x}_{\bf i}&=& \frac{1}{2}(d^{\dagger}_{\bf i} f^{\;}_{\bf i} + f^{\dagger}_{\bf i} d^{\;}_{\bf i})
\nonumber \\
\tau^{y}_{\bf i}&=& \frac{i}{2}(f^{\dagger}_{\bf i} d^{\;}_{\bf i} - d^{\dagger}_{\bf i} f^{\;}_{\bf i})
\nonumber \\
\tau^{z}_{\bf i}&=& \frac{1}{2} (n^d_{\bf i}-n^f_{\bf i}).
\end{eqnarray}
The expression for $H$ in the new language is:
\begin{eqnarray}
H &=& e_d \sum_{{\bf i}, \sigma} n_{i \sigma} +
\sum_{\bf \langle i, j \rangle, \sigma} t_{\sigma} 
(c^{\dagger}_{{\bf i} \sigma} c^{\;}_{{\bf j} \sigma} +  
c^{\dagger}_{{\bf j} \sigma} c^{\;}_{{\bf i} \sigma})
\nonumber \\
&+& U^{fd} \sum_{\bf i} n_{{\bf i} \uparrow} n_{{\bf i} \downarrow}
+ B_z \sum_{\bf i} \tau^{z}_i,
\label{Hubbard}
\end{eqnarray}
where $e_d=\frac {1}{2} (\epsilon_d + \epsilon_f)$ and 
$B_z=\epsilon_d - \epsilon_f$. The new version of $H$ is
a Hubbard model with different hoppings for each spin flavor,
$t_{\uparrow}= t_d$ and $t_{\downarrow} = t_f$, plus a Zeeman coupling
with a magnetic field $B_z$.
Both terms break the $SU(2)$ symmetry of the original Hubbard 
model ($t_{\uparrow}=t_{\downarrow}$ and $B_z=0$). The remaining
symmetries are the $U(1)$ groups associated with the conservation of
the total charge and the total $\tau^z$. 
In the original language, these $U(1)$ symmetries correspond to the 
conservation of the total number of particles in each band 
($[H,\sum_{\bf i} n^f_{\bf i}]=[H,\sum_{\bf i} n^d_{\bf i}]=0$).

I will consider from now on the half filled case, i.e. one particle 
per site. For this concentration, it is well known that the Hubbard model 
in the strong coupling limit
can be reduced to an effective Heisenberg model.
In a similar fashion, I can reduce $H$ to an effective spin model when 
$t_{\sigma} \ll U^{fd}$. The lowest energy subspace for infinite $U^{fd}$ is 
the one generated by states having one particle at each site, i.e., 
the charge degrees of freedom are frozen (the system is a Mott insulator)
and an effective spin is localized at each site.
In this limit there is a complete spin degeneracy  because the energy
does not depend on the orientation of each spin. 
To lift this degeneracy it is necessary to consider the lowest order 
processes in $t_{\sigma}/U^{fd}$.  This can be done by a canonical transformation 
which eliminates the linear terms in the hopping $t_{\sigma}$ 
and keeps the terms of quadratic order. Up to an 
irrelevant constant $C=-N{\cal Z}J_z/8$, where 
${\cal Z}$ is the coordination number and $N$ is the number of sites, 
the resulting effective spin Hamiltonian is \cite{Fath}:
\begin{equation}
H_{eff}= \sum_{\bf \langle i,j \rangle} J_{z} \tau^{z}_{\bf i} \tau^{z}_{\bf j}
+ J_{\perp} (\tau^{x}_{\bf i} \tau^{x}_{\bf j} + \tau^{y}_{\bf i} \tau^{y}_{\bf j})
+ B_z \sum_{\bf i} \tau^{z}_{\bf i},
\end{equation}
with $J_z = \frac {2(t^2_{\uparrow}+t^2_{\downarrow})}{U^{fd}}$ and 
$J_{\perp} = \frac {4 t_{\uparrow}t_{\downarrow}}{U^{fd}}$.
$H_{eff}$ is a spin 1/2 $xxz$ model with an applied magnetic 
field along the ${\bf \hat z}$ direction. 
The model is Ising-like ($J_z > J_{\perp}$). 
However it is important to consider the whole phase diagram 
because the ratio $J_z/J_{\perp}$ can take any value if non-zero nearest-neighbor 
repulsions are added to $H$. 

$H_{eff}$ has been exactly solved in one dimension by means of the 
Bethe anzatz technique \cite{Yang}. The one dimensional (1D) 
quantum phase diagram is similar to the 2D one that
I describe below. The only important difference is that,
as required by the Mermin-Wagner theorem \cite{Mermin},
the excitonic condensate is critical at zero temperature 
(power law correlations) for the 1D case.  
The phase diagram of $H_{eff}$ has  
been determined recently for 2D
systems by solving up to $96 \times 96$
lattices with quantum Monte Carlo loop algorithm \cite{Schmid}.
The zero temperature phase diagram is shown in Fig.~1 and
the corresponding name of each phase translated back to the 
original language of the FKM. Since $H_{eff}$ is symmetric under a reflection 
in the $xy$ plane, the phase diagram must be symmetric under a change 
of sign of the magnetic field. 
The fully polarized solutions, obtained for 
large values of  $|B_z|$, correspond to a full $f$ 
band for positive $B_z$ ($\epsilon_f \ll \epsilon_d$)
and a full $d$ band for $B_z$ negative ($\epsilon_f \gg \epsilon_d$).
The spectrum of both phases has a finite charge transfer (pseudospin
gap)  $\Delta_{CT}=|B_z-{\cal Z}(|J_{\perp}|+J_z)/2|$.
This gap vanishes at the quantum critical points 
$|B_{z}^{c}|=\frac {\cal Z}{2}(|J_{\perp}|+J_z)$, which are the boundaries for the
mixed valence phase that emerges when the $f$ and $d$ bands are 
sufficiently close: $|\epsilon_f-\epsilon_d|< \frac {\cal Z}{2}(|J_{\perp}|+J_z)$. 
If $J_z >J_{\perp}$, two phases are possible within the mixed 
valence regime. For small values of $|B_z|$, the $J_z$ term dominates and
induces a longitudinal antiferromagnetic (AFM) phase (chessboard state in the 
original language). When $|B_z|$ is larger than a critical value, 
the magnetic field suppresses the Ising-like ordering and the $J_{\perp}$ 
term induces a magnetic ordered state in the $xy$ plane(BEC of electron-hole pairs). 
The line separating the orbitally ordered state and
the BEC corresponds to a first order transition. 
This line ends at the Heisenberg point ($J_z=J_{\perp}$) where both 
phases coexist. For $J_z < J_{\perp}$, 
the only phase in the mixed valence regime is the BEC.
In a real material, changing $B_z=\epsilon_d-\epsilon_f$ can be achieved 
by applying pressure or alloying.
\begin{figure}[htb]
\includegraphics[angle=-90,width=8.0cm,scale=1.0]{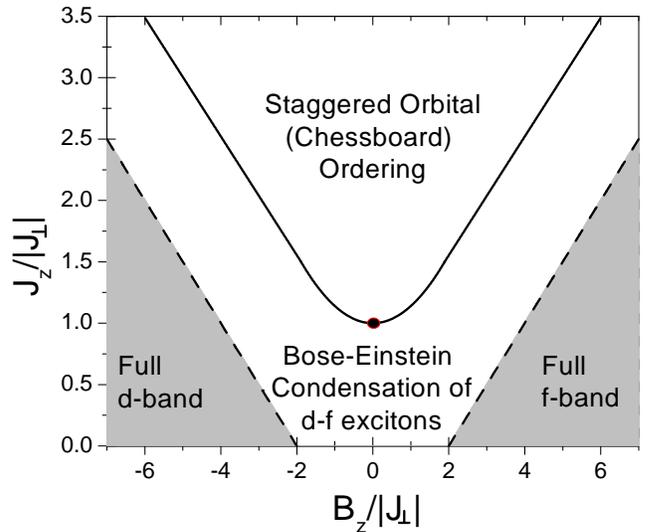}
\vspace{-1cm}
\caption{Two dimensional quantum phase diagram of $H_{eff}$ obtained 
from Ref.~\cite{Schmid}. The small circle indicates 
the position of the Heisenberg point. The dashed line denotes 
the quantum phase transition between the mixed valence (non-shadowed) and 
the non-mixed valence regimes.}
\label{fig1}
\end{figure}

The $d-f$ exciton condensate has a built-in electric polarization \cite{Portengen}.
This can be easily seen by realizing that the spin version of the 
order parameter for the BEC is the uniform $xy$ magnetization for negative 
$J_{\perp}$ (condensation at ${\bf k}={\bf 0}$) and the staggered $xy$ magnetization for
positive $J_{\perp}$ 
(condensation at the AFM wave vector ${\bf k = Q}$):
\begin{eqnarray}
{\bf M}^{\perp} &=& \sum_{\bf i} (\tau^{x}_{\bf i} \; {\bf \hat x} + 
 \tau^{y}_{\bf i} \; {\bf \hat y})
\;\;\;\;\;\;\;\;\;\;\;\; for \;\;\; J_{\perp}<0
\nonumber \\
{\bf M}^{\perp}_{ST}&=& \sum_{\bf i}  e^{i {\bf Q \cdot r_i}} (\tau^{x}_{\bf i} 
\; {\bf \hat x} + \tau^{y}_{\bf i} \; {\bf \hat y} )
\;\;\; for \;\;\; J_{\perp}>0.
\end{eqnarray}
Since ${\bf M}^{\perp}$ is a two dimensional vector,
it can also be represented by 
a complex number $|{\bf M}^{\perp}| e^{i \phi}$
($\tan{\phi}=M^{x}/M^{y}$) which is 
the usual expression for the order parameter of the BEC. 
The sign of $J_{\perp}$ is determined
by the relative sign of $t_d$ and $t_f$. On the other 
hand, the uniform polarization operator is \cite{Portengen}: 
\begin{equation}
{\bf P}= \frac {\boldsymbol{\mu}}{\Omega} \sum_{\bf i} (d^{\dagger}_{\bf i} f^{\;}_{\bf i}+
d^{\;}_{\bf i} f^{\dagger}_{\bf i})= \frac {2 \boldsymbol{\mu}}{\Omega} M^{x},
\end{equation}
where $\boldsymbol{\mu}$ is the inter-band dipole matrix element and $\Omega$
is the volume of the system. Therefore, the condensate has a built-in electric 
polarization which is proportional to the ${\bf \hat x}$ (real) component of 
its order parameter. For positive  $J_{\perp}$, the condensate becomes 
AFE because the staggered electric polarization:
\begin{equation}
{\bf P}_{ST}= \frac {\boldsymbol{\mu}}{\Omega} \sum_{\bf i} 
e^{i {\bf Q \cdot r_i}} 
(d^{\dagger}_{\bf i} f^{\;}_{\bf i}+ d^{\;}_{\bf i} f^{\dagger}_{\bf i})
= \frac {2 {\boldsymbol{\mu}}}{\Omega} M^{x}_{ST}
\end{equation}
is proportional to $M^{x}_{ST}$.

The three dimensional (3D) quantum phase diagram of $H_{eff}$ \cite{Fisher} is similar to the 
two dimensional (2D) one shown in Fig.1. The same 
is not true for the finite temperature phase diagrams due
to the Mermin-Wagner theorem \cite{Mermin}. The transition 
temperature associated to the BEC  
is finite only for the 3D case. From the 
finite temperature phase diagram obtained in Ref.~\cite{Schmid},
the BEC of electron-hole pairs undergoes a  Kosterlitz-Thouless phase transition 
in a 2D system. 

I will now analyze the effect that a time dependent electric field
${\bf E} e^{i \omega t}$ induces in our FE or AFE condensate of excitons. 
The coupling term between the electric field and the uniform polarization:
\begin{equation}
H_I= {\bf E \cdot \boldsymbol{\mu}} \; \frac{e^{i \omega t}} {\Omega}  \sum_{\bf i} 
(d^{\dagger}_{\bf i} f^{\;}_{\bf i}+ d^{\;}_{\bf i} f^{\dagger}_{\bf i}),
\label{ef}
\end{equation}
corresponds, in the spin language, to the application of a uniform time 
dependent magnetic field 
${\bf B}_1(t)= 2 {\bf E \cdot \boldsymbol{\mu}}\;e^{i \omega t}  {\bf \hat x}/\Omega$. 
From the point of view of the spin variables this is like
a magnetic resonance experiment
since $H_{eff}$
already includes a uniform static field $B_z {\bf \hat z}$. 
Therefore, the equivalent magnetic system 
will have a resonant absorption at the frequency which tends to
$\omega_0= B_z/\hbar$ when ${\bf B}_1$ tends to zero. Back to the original language, 
this means that for small electric fields the optical 
absorption will be resonant at $\hbar \omega_0=\epsilon_d-\epsilon_f$. 
This is an experimental fingerprint of the excitonic condensate. 

Since the above theory is only valid in the strong coupling limit, 
it is natural to ask whether the chessboard ordering and the excitonic 
condensate survive in the intermediate and weak coupling regimes. 
To answer this question it is more convenient to use the 
Hubbard-like representation of $H$ (see Eq.~(\ref{Hubbard})). The 
dispersion relation for the non-interacting part of $H$ 
($U^{fd}=0$) is: 
$\epsilon({\bf k},\sigma)= e_d + B_z \sigma + 2t_{\sigma} \sum_{\nu} \cos(k_{\nu})$.
At half filling and for $B_z=0$ the Fermi surface of the non-interacting 
problem nests at ${\bf k=Q}$. This indicates
that an infinitesimal value of $U^{fd}$ is sufficient to induce an AFM   
(chessboard ordering in the original language) instability. Again the presence 
of a non-zero magnetic field will induce a transition from the orbitally 
ordered state to the BEC of excitons. For the 3D case, 
if $U^{fd}< U^{fd}_{c}\sim 2.85 (|t_a|+|t_b|)$ the magnetic field $B_z$ induces 
an insulator-metal transition before the saturation of the magnetization (non-mixed valence regime)
is reached \cite{Dogen}.
This means that for weak coupling a new metallic phase appears between the BEC of exctions 
and the non-mixed valence regime (see Fig.1).
For these reasons, I expect the phase diagram of $H$ to contain 
an electric polarized BEC in the weak and intermediate coupling regimes as well.
This subject will be studied more extensively in  Ref.~\cite{Jim}.

What happens if we consider electrons instead of spinless fermions?
In this case each orbital can be occupied by two electrons 
and therefore it is natural to include local Coulomb 
repulsions $U^{ff}$ and  $U^{dd}$. By doing so the FKM is replaced by a two orbital
Hubbard model and the large $U^{\alpha,\beta}$ expansion gives 
rise to a Kugel-Khomskii like model \cite{Kugel} containing spin $\bf s$ (magnetic) and 
pseudospin $\boldsymbol{\tau}$ (orbital) degrees of freedom. If $t_a=-t_b=t$ and $U^{ff}=U^{dd}=U$
(the most general case will be analyzed in Ref.~\cite{Cristian}), the effective spin Hamiltonian is:
\begin{eqnarray}
H^{s,\tau}_{eff}&=&\frac{J_0}{2} \sum_{\langle \bf i,j \rangle} {\bf S_i}\cdot {\bf S_j} 
+(J_z-J_0) \sum_{\langle \bf i,j \rangle} \tau^{z}_{\bf i} \tau^{z}_{\bf j}
+ B_z \sum_{\bf i} \tau^z_{\bf i}
\nonumber \\
&+& 2 \sum_{\langle \bf i,j \rangle} [ {J_0} {\bf \tau^z_i}  {\bf \tau^z_j} 
- J_{\perp} (\tau^{x}_{\bf i} \tau^{x}_{\bf j}+\tau^{y}_{\bf i} \tau^{y}_{\bf j})] 
({\bf S_i}\cdot {\bf S_j}+ \frac{1}{4}),
\nonumber
\label{KK}
\end{eqnarray}
where $J_0=4t^2/U$. The ground state of $H^{s,\tau}_{eff}$ is  ferromagnetic (FM) if 
$J_0<J^{c}_0$. For the FM solution, $H^{s,\tau}_{eff}$ reduces to $H_{eff}$ because all the 
electrons have the same spin orientation and hence can be considered as spinless fermions.
Therefore the charge degrees of freedom ($\boldsymbol{\tau}$) of the FM solution are exactly 
described by $H_{eff}$, and the phase diagram is the one of Fig.~1; i.e., ferromagnetism 
coexists with chessboard ordering or a FE BEC 
of excitons. If $J_0>J^{c}_0$, the system becomes AFM. In this case the 
effective transverse coupling for the pseudospin variables, 
$J^{eff}_{\perp}= -2 J_{\perp} \langle ({\bf S_i}\cdot {\bf S_j}+ \frac{1}{4}) \rangle$,
changes its sign because $\langle ({\bf S_i} \cdot {\bf S_j}+ \frac{1}{4}) \rangle$ turns 
to be negative. For this reason the AFM solution coexists with an AFE condensate.
This means that when a magnetic field induces a transition from an AFM phase to a FM one,
it simultaneously changes the electric polarization from AFE to FE. This important 
property gives rise to new technological applications.

From now on I will continue with the spinless case just to isolate the basic 
mechanism for FE and AFE which is associated with the charge degrees of freedom. 
An important aspect of this analysis is the inclusion of a non-zero 
hybridization term:
\begin{equation}
H_{V}= V \sum_{{\bf i},\nu} 
(d^{\dagger}_{\bf i} f^{\;}_{\bf i+ {\hat e}_{\nu}} +
f^{\dagger}_{\bf i+ {\hat e}_{\nu}} d^{\;}_{\bf i} -
f^{\dagger}_{\bf i} d^{\;}_{\bf i+ {\hat e}_{\nu}} -
d^{\dagger}_{\bf i+ {\hat e}_{\nu}} f^{\;}_{\bf i}),
\end{equation}
where $\nu$ runs over the different spatial directions ($x$, $y$ and $z$ in three 
dimensions). The different signs in the hybridization terms are due to the 
different parities of the two orbitals.  
(The crystal has inversion symmetry.) By adding  
$H_V$ to $H$, we get the following additional terms for $H_{eff}$ (large $U$ expansion):
\begin{eqnarray}
H'_{eff} &=& J' \sum_{{\bf i},\nu} {\boldsymbol{\tau}_i} \cdot {\boldsymbol{\tau}_{i+{\hat e}_{\nu}}}
+J'_{xz} \sum_{{\bf i},\nu} (\tau^z_{\bf i}\tau^x_{\bf i+{\hat e}_{\nu}}
- \tau^z_{\bf i+{\hat e}_{\nu}} \tau^x_{\bf i})
\nonumber \\
&-&  2J' \sum_{{\bf i},\nu}  \tau^x_{\bf i} \tau^x_{\bf i+{\hat e}_{\nu}},
\nonumber
\end{eqnarray}
where $J'=4V^2/U^{fd}$ and $J'_{xz}=4V(t_a+t_b)/U^{fd}$. Let me now consider 
the perturbative effects of the hybridization ($J',J'_{xz} \ll J_z,J_{\perp}$) on the phase diagram 
of Fig.~1 for the FE case ($J_{\perp}<0$). The first term is a Heisenberg interaction 
which just produces a renormalization of $J_z$ and $J_{\perp}$.
The mean value of the second  term is zero in 
the FE phase (BEC of excitons) and therefore they do not make any contribution. The last
term introduces an easy axis anisotropy along the ${\bf \hat x}$ direction and lifts the 
U(1) degeneracy. The BEC then is replaced by an Ising-like FE state characterized by the
breaking of the remaining $Z_2$ symmetry. Therefore, the spontaneous 
ferroelecricity remains when the hybridization is included perturbatively;
however, the resonant response to a time dependent electric field (see Eq.~[\ref{ef}])
disappears due to the absence of Goldstone modes. In other words, the hybridization makes the 
electronically induced FE phase similar to the ones induced by structural phase transitions.

In summary, I derived the phase diagram of the FKM with a $t_{f}$ hopping term
in the strong coupling limit. The insulating phase obtained at half filling
has a transition from a non-mixed valence to a mixed valence regime as a function of the energy 
difference between the centers of both bands. Two different phases are present in the 
mixed valence regime: a BEC of excitons with a built-in electrical polarization
which starts just at the valence transition and an orbitally ordered (chessboard) state which appears 
when the centers of the bands are sufficiently close. These results were extended to the intermediate 
and weak coupling regimes due to the nesting property of the Fermi surface of hypercubic lattices.
I also mentioned the effect of including spin degrees of freedom in $H$: the interplay 
between magnetic and charge degrees of freedom gives rise to the coexistence of FE 
and FM phases which are coupled to each other. This 
opens the possibility of controlling optical properties by applying magnetic fields
or controlling magnetic properties by applying electric fields. 

The effect of a non-zero hybridization was also considered. The main conclusion is 
that the $U(1)$ degeneracy associated to the BEC of excitons is lifted by the 
hybridization and replaced by an Ising-like FE state (broken $Z_2$ symmetry).
Then, the resonant response to a time dependent electric field disappears 
because the Goldstone modes acquire a finite mass (gap). 

These results indicate that the following characteristics are favorable to the 
formation of an electronically driven FE state: a) The system must be in a mixed-valence 
regime and the two bands involved must have different parity.
b) It is best, though not necessary, if both bands have similar bandwidths.
c) A local Coulomb repulsion ($U^{fd}$) between the different 
orbitals is required. d) The hybridization between the bands must 
be small compared to their bandwidths.  
   
I wish to thank L. Sham, J. E. Gubernatis, J. Thompson, J. Sarrao, N. Kawashima,
and A. A. Aligia for stimulating discussions.  
This work was sponsored by the US DOE under contract
W-7405-ENG-36.

\end{document}